\documentclass[prc,aps,nofootinbib,twocolumn]{revtex4}
\usepackage{epsfig}
\usepackage{graphicx}
\usepackage{amssymb}
\usepackage{hyperref}
\usepackage{color}
\usepackage{mathtools,amssymb,amsthm}

\usepackage{tikz}
\usetikzlibrary{quantikz}

\usepackage{youngtab}
\usepackage{young}
\usepackage{ytableau}

\usepackage{bm}
\begin{document}

\title{Filtering states with total spin on a quantum computer}

\author{Pooja Siwach}
\affiliation{Department of Physics, Indian Institute of Technology Roorkee, 
Roorkee-247667, Uttarakhand, India}
\author{Denis Lacroix } \email{denis.lacroix@ijclab.in2p3.fr}
\affiliation{Universit\'e Paris-Saclay, CNRS/IN2P3, IJCLab, 91405 Orsay, France}

\date{\today}
\begin{abstract}
Starting from a general wave function described on a set of spins/qubits, we propose several quantum algorithms 
to extract the components of this state on eigenstates of the total spin ${\bf S}^2$ and its azimuthal projection $S_z$. The method plays 
the role of total spin projection and gives access to the amplitudes of the initial state on a total spin basis. 
The different algorithms have various degrees of sophistication depending on 
the requested tasks. They can either solely project onto the subspace with good total spin or completely
uplift the degeneracy in this subspace. After each measurement, the state collapses to one of the spin 
eigenstates that could be used for post-processing.  For this reason, we call the method Total Quantum Spin filtering (TQSf). 
Possible applications ranging from many-body physics to random number generators are discussed.  
\end{abstract}

%\keywords{quantum computing, quantum algorithms}

%\pacs{03.65.Yz, 05.30.-d, 42.50.Lc}
\maketitle

\section{Introduction}

Given a complex problem and a set of qubits forming a quantum computer, 
what is the optimal way to encode the information on the problem in this quantum 
computer? There is certainly no unique answer to this question. A strong 
guide is the symmetry properties of the problem under consideration. Typical 
examples of interest for the present discussion are the combinatorial problems. Let us assume 
a system corresponding to a set of $n$ elements $\{ e_i \}$ with some properties. 
The answers to some questions about the system sometime do not change if we make permutations between some of the elements. We assume that the property of each element 
$e_i$ is now encoded on a qubit $| s_i \rangle$ (with $s_i\in\{0,1\}$). 
A possible basis is the basis formed by tensor product states $\{ \otimes_{i} | s_i \rangle \}$, which we call the natural basis (NB). In this basis, 
the system can be represented on a quantum computer 
by a wave function
\begin{eqnarray}
| \Psi \rangle = \sum_{s_i\in\{0,1\}} \Psi_{s_1, \cdots, s_N} | s_1, \cdots, s_n \rangle. \label{eq:general}
\end{eqnarray}
The invariance of the wave function will then reflect the invariance of the combinatorial problem with the permutation of some of the elements with respect to the exchange of spins. The direct consequence of the invariance, which is well known in physics,  are (i) that specific recombinations of the natural states will show up in Eq.~\eqref{eq:general};  (ii) that the problem might 
be more efficiently treated by considering the proper combination of states prior to the encoding of the problem. The latter is the underlying idea of permutational quantum computing (PQC) introduced in Refs.~\cite{Mar02,Mar05} and further discussed in Ref.~\cite{Jor10}. 

The PQC technique uses an alternative basis for the quantum processor unit (QPU) connected to the eigenstates of the total spin and its azimuthal component. The interest in such a basis for combinatorial problems is not surprising. It was indeed realized in the early times of quantum mechanics that these states are intimately linked to the permutation symmetry group $S_n$ (for a nice historical overview, we recommend the Ref.~\cite{Low70}). Therefore, this basis is widely used to solve quantum many-body problems using the total spin algebra. Its links to the permutation group are well documented in many textbooks, to quote some of them \cite{Mes62,Kap75,Ham12}. 

In the following, we will simply use the terminology ``Total Spin Basis" (TSB) for the basis to be used in the PQC framework. Finding the 
TSB is equivalent to construct the complete set of the irreducible representations of the symmetric group. The construction of these representations from the natural basis on a quantum computer has attracted a lot of attention primarily due to its usefulness in quantum many-body problems appearing in physics and chemistry \cite{McA20}. For example, an efficient quantum algorithm based on the Schur transformation was proposed in Ref.~\cite{Bac06} (see also Refs.~\cite{Kir17,Kir18,Kro19, Hav18}). We note that  a classical algorithm was proposed in Ref.~\cite{Hav19} that can compete in computing the amplitudes in the PQC.   

The possibility of preparing and using states of the TSB on a quantum computer is also of great interest for studying interacting particles with spins when the total spin commutes with the Hamiltonian. In the classical simulations, the use of such symmetry 
automatically gives a focus on the relevant subspace of the Hilbert space. Significant efforts are being made currently to prepare many-body states in quantum computers that automatically preserve the spin symmetry \cite{Sug16,Sug19,Lui19,Gar20,Sek20}, 
intending to obtain more optimal states that can be used in variational calculations. 

In the present work, we have a different objective. Given an initial state that is not necessarily 
an eigenstate of the total spin or of the azimuthal projection of the total spin, we propose a general algorithm to compute the amplitudes of the 
state on the TSB states. It turns out that this algorithm can also be used (i) to select a specific component of the initial state projected 
on good total and azimuthal spin, playing the role of spin projection, or (ii) to obtain specific states of the TSB. For this reason, we call it 
the Total Quantum Spin filtering (TQSf) method.  We note that the objective is directly related to the symmetry breaking/symmetry restoration problem, and discussion on its formulation on quantum computers can be found in Refs.~\cite{Whi13,Mol16,Rya18,Chi19,Tsu20,Lac20}.
 
%%% to be continued. 
\section{Quantum algorithms for the TQSf method - (Method and notation)}

We consider here an ensemble of $n$ spins labelled by $i$ with components up or down denoted by $\{ | \sigma \rangle_i = | \pm\rangle_i \}_{i=0,{n-1}}$.  We use the convention $| 0_i \rangle = |+\rangle_i$ and  $|1_i \rangle = | -\rangle_i$ to match with the standard notations in quantum computing. The total spin operator of the system is defined as ${\bf S} = \sum_i {\bf S}_i$, where ${\bf S}_i$ denotes the spin operator associated with the particle $i$ that is linked to the standard Pauli matrices through ${\bf S}_i = \frac{1}{2} (X_i, Y_i, Z_i)$. These three operators are completed by the identity operator $I_i$. 

We consider a general wave function $| \Psi \rangle$ given in the natural basis by Eq.~\eqref{eq:general}. We know that the eigenstates 
of the commuting variables ${\bf S}^2$ and ${S}_z$ form a complete basis for the Hilbert space of $n$ qubits.  The possible eigenvalues of ${\bf S}^2$ and ${S}_z$ are given by $S(S+1)$ and $M$ (assuming $\hbar=1$) respectively, with the constraints $S \le n/2$ and $-S \le  M \le +S$.

We introduce the set of projectors ${\cal P}_{[S,M]}$ that projects on the sub-space associated with the eigenvalues $(S,M)$.  
Our first objective is to obtain the amplitudes of the initial state decomposition $A_{S,M} \equiv \langle \Psi |{\cal P}_{[S,M]}| \Psi \rangle$
and eventually extracts one of the projected normalized states given by
\begin{eqnarray}
| \Psi_{S,M} \rangle = A^{-1/2}_{S,M} {\cal P}_{[S,M]}| \Psi \rangle. \nonumber
\end{eqnarray}

To achieve this objective, we apply the technique proposed in Ref.~\cite{Lac20}. We consider two separate operators $U_S$ and $U_z$, allowing for the discrimination of ${\bf S}^2$ and ${S}_z$ when used in the Quantum Phase Estimation (QPE) algorithm \cite{Nie02,Fan19,Ovr03,Ovr07}. As a result, the projection on the states $| \Psi_{S,M} \rangle $ is automatic when the ancillary qubits used in the QPE are measured.

The operators used to discriminate the different components are taken as $U_{S/z} = e^{2\pi i \alpha_{S/z}(n) O_{S/z}}$, where $O_S$ 
and $O_z$ are operators with known eigenvalues. The eigenvalues, denoted by $\{ \lambda^S_i \}$ and $\{ \lambda^z_i \}$, are 
proportional to those of ${\bf S}^2$ and $S_z$, respectively. Furthermore, $\alpha_{S}(n)$ and $\alpha_{z}(n)$ should be chosen in a very specific way. 
These parameters should ensure that, for all eigenvalues, the quantities $\alpha_{x}(n)\lambda^x_i$  verifies 
$0 \le  \alpha_{x}(n)\lambda^x_i < 1$ and that these quantities always correspond to a binary fraction with a finite number of terms. Moreover, denoting the number of extra ancillary qubits used in the QPE, by $n_S$ and $n_z$ respectively for $U_S$ and $U_z$, these numbers should be chosen as the minimal values such that $2^{n_x} \alpha_{x}(n)\lambda^x_i$ are positive integers for all eigenvalues. 
    
There is some flexibility in the choice of both $U_S$ and $U_z$. First, we consider the total $S_z$ component. This component verifies:
\begin{eqnarray}
S_z &=&  N_0 - N_1 ~{\rm with} ~N_0+N_1 = n I ,\nonumber 
\end{eqnarray}   
where $N_0=\frac{1}{2} \sum_{k}( I_k + Z_k)$ (resp. $N_1=\frac{1}{2} \sum_{k} (I_k - Z_k)$) is the  operator that counts the number of $0$ (resp. the number of $1$ in the state).  $S_z$, $N_0$, and $N_1$ are commuting operators, and the states of the natural basis are eigenstates of these operators. To select the states with good particle number or, equivalently, eigenstates of $S_z$, we use the QPE applied on $N_1$. With the constraint listed above, a convenient choice is 
\begin{eqnarray}
U_z = \exp\left\{ 2\pi i \frac{N_1}{2^{n_z}}\right\}.
\end{eqnarray}
The eigenvalues of $N_1$ range from 0 to $n$. Accordingly, the minimal possible value for $n_z$ is such that $n_z > \ln n/\ln 2$.  With this choice, the filtering of states with respect to the eigenvalues of $S_z$ becomes strictly equivalent to the particle number projection illustrated in Ref.~\cite{Lac20}. Therefore, in the natural basis, $U_z$ is given by a product of phase operators
\begin{eqnarray}
\widetilde{U}_z &=&  \prod_k \left[ \begin{array}{cc} 1 & 0 
\\ 0&  e^{i\pi/2^{n_z - 1}} \end{array}\right]_k.  \nonumber
\end{eqnarray}    

We now consider the projection on total spin ${\bf S^2}$. For $n$ qubits, the eigenvalues of this operator are positive and verifies $\lambda_S \le n(n+2)/4$. Depending on the fact that $n$ is even or odd, we propose the following form of $U^{e/o}_S$:
\begin{eqnarray}
U^{\rm e}_S = \exp \left\{  2\pi i\frac{{\bf S}^2}{2^{n_S+1} }  \right\} , 
~U^{\rm o}_S = \exp \left\{  2\pi i\frac{({\bf S}^2 - 3/4)}{2^{n_S} }  \right\}. 
\label{eq:usevenodd}
\end{eqnarray} 
The number of ancillary qubits has the constraints
%\begin{eqnarray}
%n_S \ge \left\lceil\frac{\ln k(k+1)}{\ln 2}\right\rceil ~{\rm (even)}, ~n_S\ge \left\lceil\frac{\ln k(k+2)}{\ln 2}\right\rceil ~{\rm (odd)},
%\label{eq:nsoddeven}
%\end{eqnarray}
\begin{eqnarray}
	n_S>  \frac{\ln k(k+1)}{\ln 2}-1 ~{\rm (even)}, ~~n_S>\frac{\ln k(k+2)}{\ln 2} ~{\rm (odd)},
	\label{eq:nsoddeven}
\end{eqnarray}
respectively for even $n=2k$ and odd $n=2k+1$ cases. 

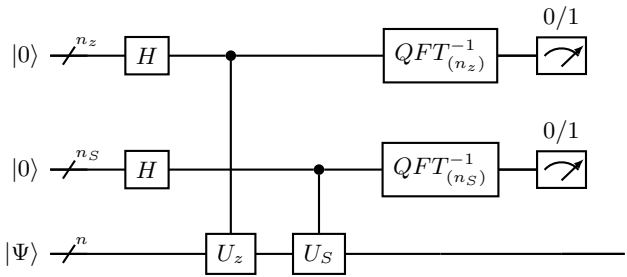
\begin{figure}
 \begin{quantikz}
 \lstick{\ket{0}}  &  \push{ \qwbundle{n_z}} & \gate{H}                         &  \ctrl{2} & \push{}  & \gate{QFT^{-1}_{(n_z)}} &\meter{0/1}  \qw \\% \push{}   \qw \\
 %\lstick{\ket{0}}  &  \gate{H}                         &   \ctrl{2} &  \push{}  &\push{} & & \meter{0/1} \qw \\ % \ctrl{3}  \qw \\
 \lstick{\ket{0}}  &  \push{ \qwbundle{n_S}} & \gate{H}                            &   \push{}  & \ctrl{1} & \gate{QFT^{-1}_{(n_S)}}& \meter{0/1}  \qw \\ %  \push{}   \qw \\
 \lstick{\ket{\Psi}}  & \push{ \qwbundle{n}} &\push{} &  \gate{U_z}   &  \gate{U_S}  &   \push{}  & \push{} & \push{} %\ket{\Psi_{S,M}}
 \qw  %  \targ{}   \qw
\end{quantikz}
 \caption{Schematic illustration of the  circuit used in the present work to filter the states using the total spin ${\bf S}^2$ and $S_z$ 
 components.}
\label{fig:circuit_s} 
\end{figure}

In practice, to compute the $U^{e/o}_S$ operators, we use the standard formula \cite{Low70}:
\begin{eqnarray}
{\bf S}^2 &=&  \frac{n (4-n)}{4} I +  \sum_{i<j, j=0}^{n-1}  P_{ij}, \label{eq:transpos}
\end{eqnarray}
that generalizes the Dirac identity originally derived for two spins in Ref.~\cite{Dir35}.   The set of operators 
$P_{ij}$ are the transposition operators given by
\begin{eqnarray}
P_{ij} &=& \frac{1}{2} (I+X_iX_j +Y_iY_j +Z_iZ_j) \nonumber.
\end{eqnarray}
We have in particular $P_{ij}| \delta_i \delta_j\rangle = | \delta_j \delta_i \rangle$ and $P^2_{ij}=P_{ij}$. With 
the formula given in Eq.~\eqref{eq:transpos}, the link between total spin operator and permutation 
group becomes explicit. Some aspects of transpositions and their use in directly constructing states with good total spins were discussed in Ref.~\cite{Sek20}. 
In the quantum computing context, the transposition operators are nothing but the 
SWAP operators. In the present work, we implement the $U_S$ operators [Eq.~\eqref{eq:usevenodd}] using 
the Trotter-Suzuki decomposition technique \cite{Tro59,McA20} based on the expression given by Eq.~\eqref{eq:transpos} and by noting that:
\begin{eqnarray}
e^{i\alpha P_{ij}} = \cos{\alpha} I + i P_{ij} \sin(\alpha).
\end{eqnarray} 
A schematic diagram of the circuit to perform the simultaneous selection of 
eigenstates of $S_z$ and ${\bf S}^2$ is shown in Fig.~\ref{fig:circuit_s}.  

\begin{figure*}%[h!]
\includegraphics[width=0.48\linewidth]{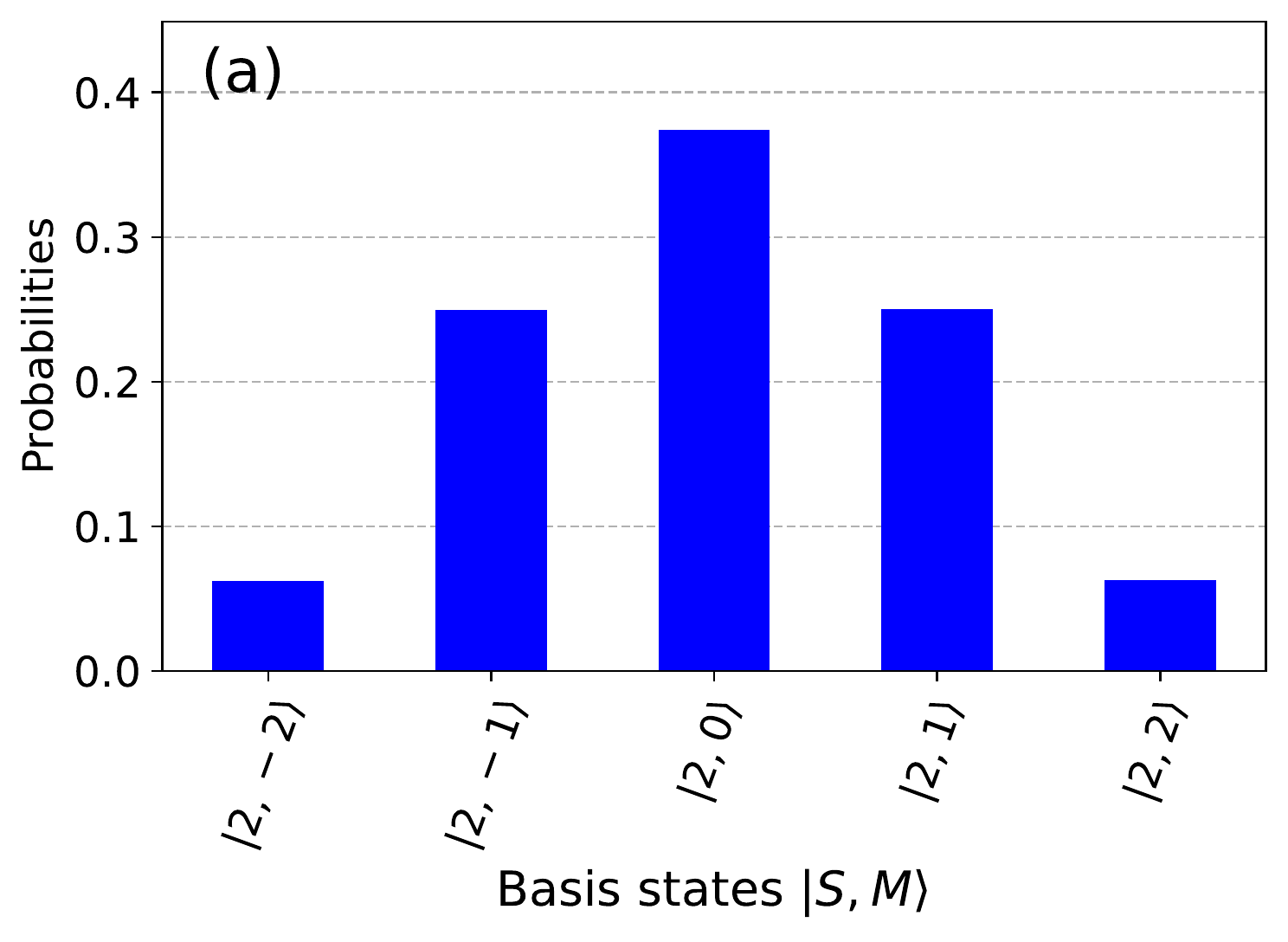}
\includegraphics[width=0.48\linewidth]{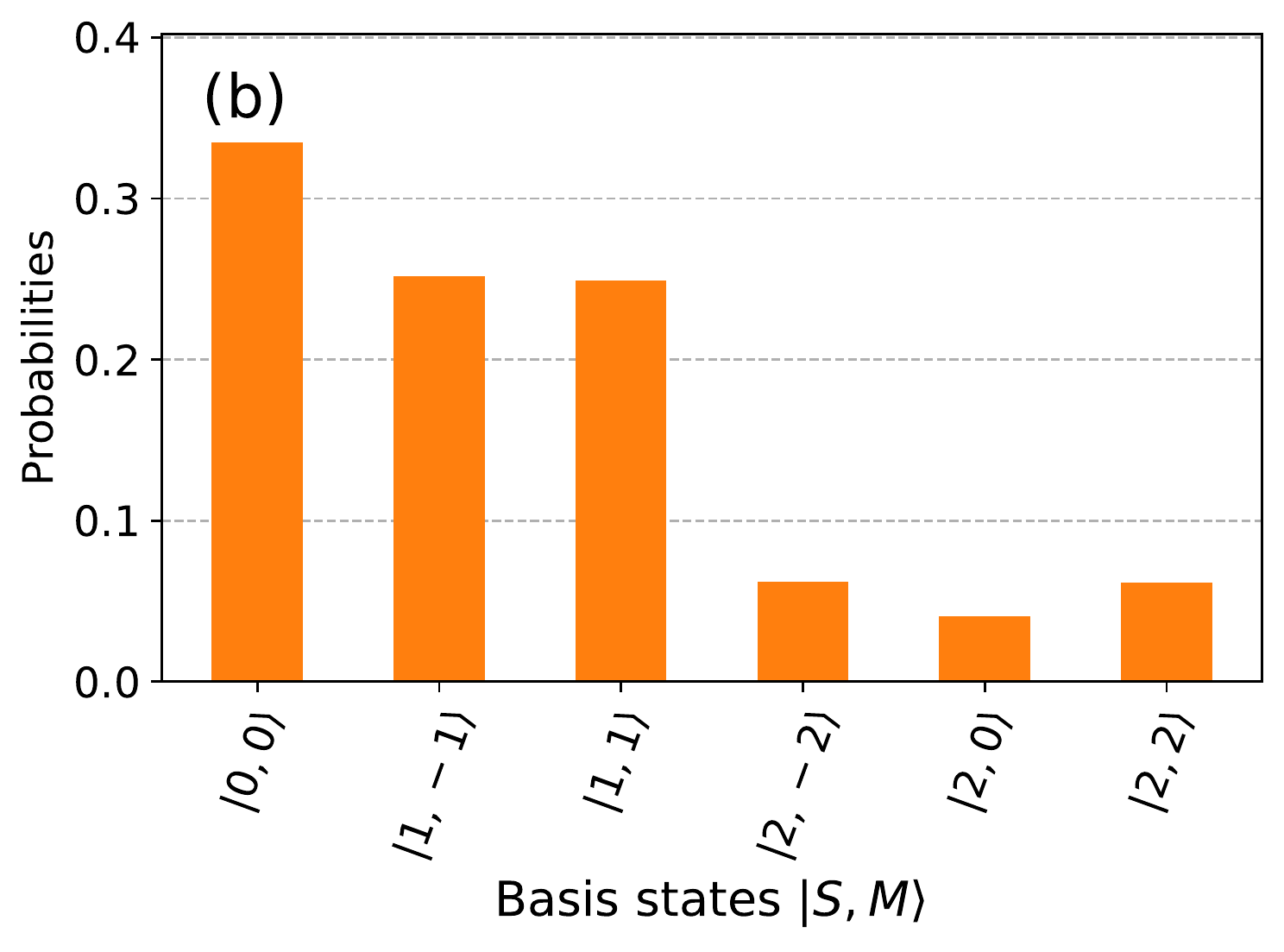}
 \caption{Illustration of the results obtained for a system described on $n=4$ qubits. In this case, the optimal choice for the number of ancillary qubits are $n_S= 2$ and $n_z =3$. Results are obtained for an initial state $|\Psi\rangle =  \prod^{3}_{k=0} H_k | 0 \rangle$ (panel (a)) and  $|\Psi\rangle = X_1 X_3 \prod^{3}_{k=0} H_k | 0 \rangle$  (panel (b)) using the qiskit software with $10^{5}$ shots of qasm simulator. 
 Here $H_k$ denotes the Hadamard gate.}
\label{fig:testsimple} 
\end{figure*}

The method proposed here is tested using the IBM toolkit qiskit \cite{Abr19}. We show in Fig.~\ref{fig:testsimple} 
the amplitudes obtained for a system described on $n=4$ qubits by measuring the ancillary qubits 
of the circuit shown in Fig.~\ref{fig:circuit_s} for two examples of initial states. For such a small number of qubits, the decomposition 
in terms of the $|\Psi_{S,M} \rangle$ can be obtained analytically. We have checked that the amplitudes obtained with the measurement 
are consistent with the analytical ones within the errors due to the finite number of measurements.      

We can further analyze the obtained results. Results displayed in panel (a) of Fig.~\ref{fig:testsimple} correspond to an initial state 
that is completely symmetric with respect to any permutation of the qubit indices. Consequently, it only decomposes on state 
$| \Psi_{S,M} \rangle$ that also has this property. Such states correspond to the states with the maximal possible eigenvalue of ${\bf S}^2$, i.e., in our case, $S=2$. In the context of group theory, the irreducible representation associated with the TSB can be represented by the Young tableau \cite{Mes62,Kap75,Ham12} with a maximum of two rows. Fully symmetric states are those represented with a single row.   
For a general initial state with $n$ qubits given by $|\Psi\rangle =  \prod^{n-1}_{k=0} H_k | 0 \rangle$, its decomposition onto the TSB will be given by
\begin{eqnarray}
|\Psi\rangle &=& \sum_{k=0}^{n} \sqrt{p_k} | \Psi_{S=n/2 , S_z = n/2 - k} \rangle,
\end{eqnarray}    
where $k$ is the eigenvalue associated to $N_1$.  For this specific initial state, the amplitudes $p_k$ are equal to $C^{k}_n/2^n$ identifying with a binomial distribution (with $p=q = 1/2$). In the large $n$ limit, this probability will tend to a Gaussian probability.  
It is interesting to mention that  a direct by-product of the approach is the possibility to generate random numbers $x_k=k/n \in [0,1]$ on an
equidistant discretized mesh according to the set of probabilities $\{ p_k \}$.
%These probabilities can eventually be changed by changing the initial state.    

We now come to the main goal of the present algorithm. When measurements are performed on the two sets of ancillary qubits respectively 
associated to $U_S$ and $U_z$, after each measurement labelled by $(\lambda)$, according to the Born rule, the total wave function 
$| \Psi^{(\lambda)}_f \rangle$ identifies with
\begin{eqnarray}
| \Psi^{(\lambda)}_f  \rangle &=& | S^{(\lambda)} \rangle \otimes  | M^{(\lambda)} \rangle \otimes | \Psi_{S^{(\lambda)} ,M^{(\lambda)}} \rangle.
\label{eq:outstate}
\end{eqnarray}     
Here, $S^{(\lambda)} $ (resp. $M^{(\lambda)}$) should be interpreted as the binary number obtained by measuring the ancillary qubits associated with $U_S$ (resp. $U_z$) in the event $\lambda$. So, after the measurement, the wave function $| \Psi^{(\lambda)}_f  \rangle$ is an eigenstate 
of both the
total spin and  its azimuthal component. Said differently, the circuit represented in Fig.~\ref{fig:circuit_s} plays the role of a funnel that 
lets only one component $(S,M)$ pass at each event, and therefore, acts as a projector on the TBS basis.  

The values $(S^{(\lambda)},  M^{(\lambda)})$ might change at each measurement unless the initial state is already an eigenstate of the total spin operators.  In general, the outcome of the circuit can be controlled solely through the initial state. Consecutively, the projected state can be used for post-processing. A direct application of the present method in physics or chemistry is to study spin systems that encounter spontaneous symmetry breaking associated with a preferred spin orientation.  If we assume that the initial state depends on a set of parameters 
$\{ \theta_i\}_{i=1,g}$, the symmetry restored state can then be used in variational approaches both prior (projection after variation) 
or after the projection (projection before variation) (see for instance Refs.~\cite{Rin80,Bla86}). 

The circuit of Fig.~\ref{fig:circuit_s} helps in achieving our first objective, which is the preparation of states with good total spin and total $z$-projection. 
This technique also works if the initial state is not fully symmetric with respect to the permutation of qubits. An example of such application is given in panel (b) of Fig.~\ref{fig:testsimple}. As we see, in this case, the state will also have components on total spins with $S < n/2$, i.e., the states corresponding to a Young tableau with two rows. There is, however, a difference between the Fully Symmetric (FS) case and the other cases. In the FS case, the Hilbert space associated with the eigenvalues $(S,M)$, denoted by ${\cal H}_{S,M}$ contains only one eigenvector of ${\bf S}^2$ and $S_z$.
In other cases, i.e., for ${\cal H}_{S,M}$ with $S<n/2$, the Hilbert space contains an ensemble of degenerated eigenstates.  For instance, in the $n=4$ qubit case, the space ${\cal H}_{0,0}$  contains two states, while ${\cal H}_{1,M}$ with $M=-1,0,1$ contains three states.  We denote by $d_{(S,M)}$ the size (degeneracy) of the ${\cal H}_{S,M}$ Hilbert space. 
In the degenerate case, the system state in Eq.~\eqref{eq:outstate} obtained after measuring the ancillary qubits will be an admixture of the different eigenstates, where the mixing coefficients will directly reflect the relative proportion of the degenerated 
states in the initial wave function.

\section{Construction of the amplitude on the complete irreducible representation by measurements}

Let us now consider a complete basis formed by eigenstates of ${\bf S}^2$ and $S_z$. We denote one element of the basis by 
$|S ,M \rangle_{g}$. The indices $g=1,d_{S,M}$ are introduced to dissociate different states belonging to the space ${\cal H}_{S,M}$. 
The system's initial state $|\Psi \rangle$ can be decomposed as
\begin{eqnarray}
|\Psi \rangle &=& \sum_{S,M} \sum_{g=1}^{d_{S,M}} c^g_{S,M} | S ,M \rangle_{g} \label{eq:irrep}.
\end{eqnarray}  
When $d_{S,M}=1$, the state $| S ,M \rangle_{1}$  will be identical with the state $| \Psi_{S ,M} \rangle$ introduced previously. Otherwise, 
$| \Psi_{S ,M} \rangle$ is an admixture of the states $| S ,M \rangle_{g}$. Here, we intend to generalize the circuit proposed in Fig.~\ref{fig:circuit_s}, in order to obtain the amplitudes $|c^g_{S,M}|^2$, and to obtain directly one of the states of the irreducible representation $| S ,M \rangle_{g}$
after the measurement of ancillary qubits.  

For this, we use the same strategy as in the PQC framework. Coming back to the Young tableaux representation, all states that are not
fully symmetric have two rows. Let us assume that these states correspond to $l_1$ and $l_2$ blocks on the first and second row, respectively (with $l_2 \le l_1$ and $l_1+l_2 = n$). 
The associated total spin corresponds to $S = (l_1 -l_2)/2$. The different $| S ,M \rangle_{g}$ have the same $(l_1,l_2)$ but differ in their symmetries with respect to the exchange of qubits. Each state can be associated to a different sequence of Young tableaux when 
including each spin/qubit one after the other \cite{Mes62,Jor10,Hav19}  (see for instance Fig.~4 of Ref.~\cite{Hav19}). The sequence of the Young tableau can be seen as an iterative procedure, where the total spin of $n$ qubits is obtained by coupling one spin at a time. 
Starting from one spin, a second spin is added and an eigenstate of the operator ${\bf S}^2_{[2]}$ is obtained. Here, the index $[2]$ indicates that the operator refers only to the first two spins. Consecutively, a third spin is coupled to find an eigenvector of  ${\bf S}^2_{[3]}$, and so on, until the 
eigenstate of ${\bf S}^2_{[n]}$ is obtained \cite{Jor10, Hav19}.  For a system with $n$ qubits exactly, ${\bf S}^2_{[n]}$ identifies  with the total spin ${\bf S}^2$ defined previously.
In the following, we denote the total spin eigenvalue for the first $k$ qubits by $S_{[k]}$, such that the eigenvalue of ${\bf S}^2_{[k]}$ is 
equal to $S_{[k]}( S_{[k]} + 1)$.   
\begin{figure}
	\begin{tikzpicture}
	\node[scale=0.75] {
 \begin{quantikz}[row sep=0.3cm,column sep=0.5cm]%
 \lstick{\ket{0}}  &  \push{ \qwbundle{n_z}} & \gate{H} & \push{}  & \push{} \cdots & \push{}  & \ctrl{4} &  \gate{QFT^{-1}} &\meter{0/1} \qw \\% \push{}   \qw \\
 %\lstick{\ket{0}}  &  \gate{H}                         &   \ctrl{2} &  \push{}  &\push{} & & \meter{0/1} \qw \\ % \ctrl{3}  \qw \\
 \lstick{\ket{0}}  &  \push{ \qwbundle{n_{[2]}}} & \gate{H} & \ctrl{3} & \push{} \cdots & \push{}   & \push{} &\gate{QFT^{-1}}& \meter{0/1} \qw \\ %  \push{}   \qw \\
 \wave&&&&&&&&&&\\
 \lstick{\ket{0}}  &  \push{ \qwbundle{n_{[n]}}} & \gate{H} & \push{} & \push{}   \cdots & \ctrl{1}   & \push{} & \gate{QFT^{-1}}& \meter{0/1} \qw \\ % 
\lstick[wires=4]{\ket{\Psi}}  & \push{} &\push{}  & \gate[wires=2][0.5cm]{U_{[2]}}& \push{} \cdots & \gate[wires=4][0.5cm]{U_{[n]}} &  \gate[wires=4][0.8cm]{U_z} & \push{} &  \push{}&  \push{}
 \qw   \\
% \lstick{\ket{\Psi}}  
 & \push{} &\push{}    &  \qw & \push{} \cdots &  &   \push{} &  \push{}  &  \push{}&  \push{}
 \qw  \\ %  \targ{}   \qw
 \wave&&&&&&&&&&\\
 %\lstick{\ket{\Psi}} 
  & \push{} &\push{}    &  \qw & \push{} \cdots &  &   \push{}  &  \push{} &  \push{}&  \push{}
 \qw  %  \targ{}   \qw
\end{quantikz}
};
\end{tikzpicture}
 \caption{Illustration of the circuit used to obtain the amplitudes $|c^g_{S,M}|^2$. After each measurement, the final state 
 of the system collapse to one of the state of the irreducible representation $| S, M \rangle_g$. }
\label{fig:circuit_sn} 
\end{figure}
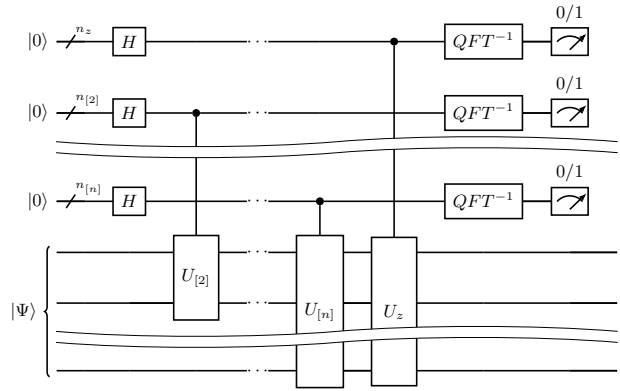

As an 
illustration, we consider the $n=4$ case used in Fig.~\ref{fig:testsimple}. The states $| 1, M \rangle$ can be generated by the three sequences of Young tableau given by
\begin{eqnarray}\label{eq:paths}
	{
		\begin{cases}
			\Yvcentermath1 {\young(1)}  \longrightarrow   \Yvcentermath1 {\young(12)} \longrightarrow   
			{\young(123)}   \longrightarrow   \Yvcentermath1 {\young(123,4)}      &   ~~{\rm path~(a)} \\
			&\\
			\Yvcentermath1 {\young(1)} \longrightarrow   \Yvcentermath1 {\young(12)}
			\longrightarrow   \Yvcentermath1 {\young(12,3)}   \longrightarrow   \Yvcentermath1 {\young(124,3)}     &    ~~{\rm path~(b)} \\
			&\\ 
			\Yvcentermath1 {\young(1)}  \longrightarrow   \Yvcentermath1 {\young(1,2)}  \longrightarrow   \Yvcentermath1 {\young(13,2)}   \longrightarrow   \Yvcentermath1 {\young(134,2)}  &    ~~{\rm path~(c)} \\
	\end{cases}}
\end{eqnarray}
Omitting $S_{[1]}$ that is always equal to $1/2$, the three  sequences in Eq.~\eqref{eq:paths} correspond to the set of eigenvalues for $[S_{[2]} \rightarrow S_{[3]} \rightarrow S_{[4]}]$
respectively given by (a) $[1  \rightarrow 3/2  \rightarrow 1]$, (b) $[1  \rightarrow 1/2  \rightarrow 1]$ and (c) $[0  \rightarrow 1/2   \rightarrow 1 ]$.

 There are several important properties to be recalled here. First, there is a one-to-one correspondence between the Young tableaux sequence and a state of the irreducible representation. Second, the state constructed by a Young tableaux sequence has a ``memory" of its path, i.e., 
 it is an eigenvalue of the full set of operators ${\bf S}^2_{[2]}, \cdots, {\bf S}^2_{[n]}$ along with the total $S_z$ components. This last property gives us a direct way to generalize the circuit given in Fig.~\ref{fig:circuit_s}  and obtain the amplitudes in Eq.~\eqref{eq:irrep}.  
 A brute force technique consists of introducing a set of ancillary qubits and perform independent QPEs for all the operators ${\bf S}^2_{[j]}$ together with the QPE associated to the total $S_z$ component. In practice, the QPE on a specific total spin ${\bf S}^2_{[j]}$  is associated 
 to a unitary operator denoted by $U_{[j]}$, which can be constructed in a similar way as the operators defined in Eq.~\eqref{eq:usevenodd} depending on whether $j$ is odd or even. The operators $U_{[j]}$ are deduced simply by replacing ${\bf S}^2$ with ${\bf S}^2_{[j]}$ and by optimizing 
 the number of ancillary qubits $n_{[j]}$ according to the accessible eigenvalues of  ${\bf S}^2_{[j]}$ as prescribed in Eq.~\eqref{eq:nsoddeven}. 
 
\begin{figure}[h!]
\includegraphics[width=0.99\linewidth]{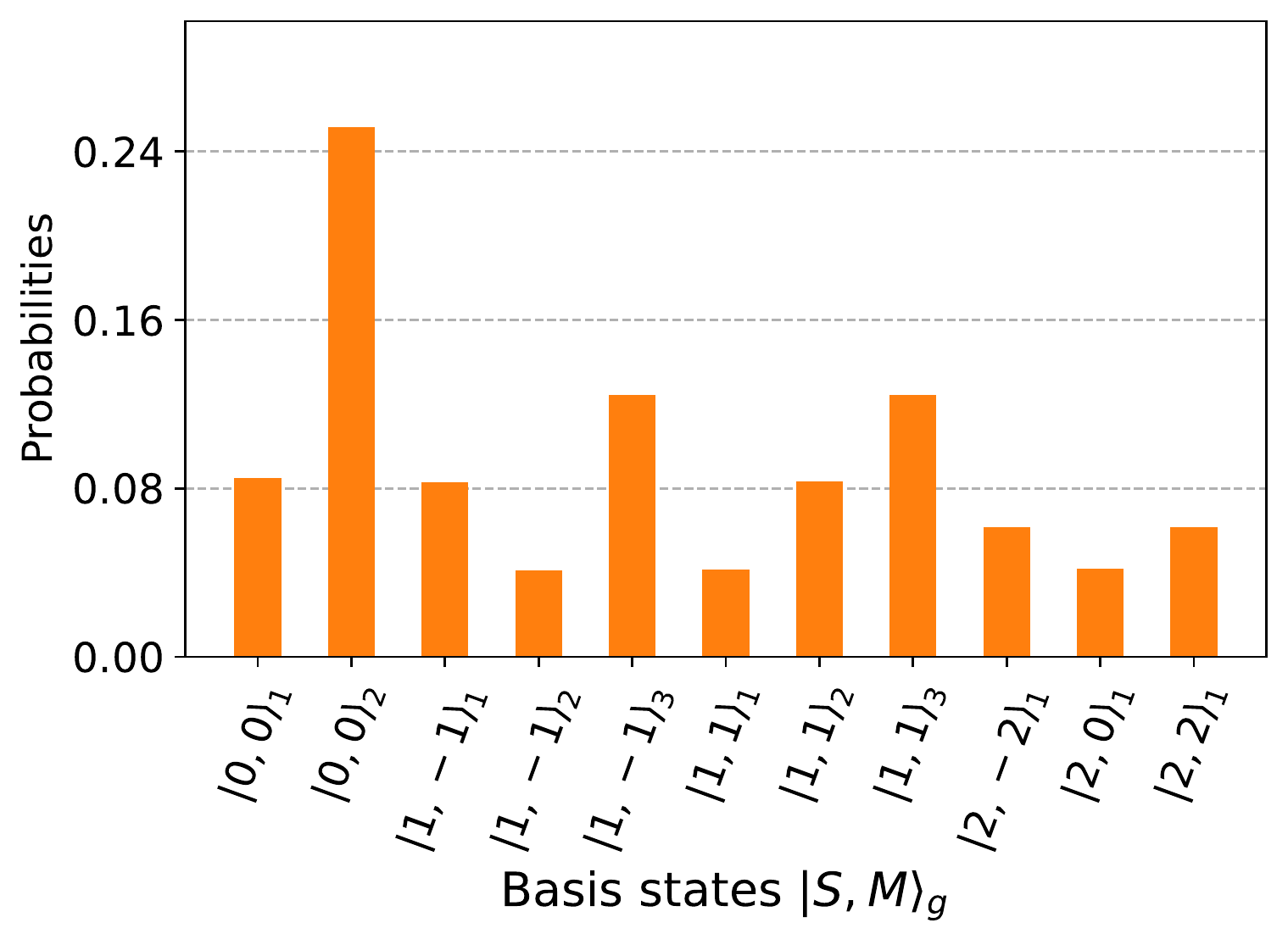}
 \caption{Results obtained for a system described on $n=4$ qubits with the same initial state as in panel (b) of Fig.~\ref{fig:testsimple} but using the circuit shown in Fig.~\ref{fig:circuit_sn}. The degeneracy in the components of states $|1,-1\rangle, |1,1\rangle$, and $|0,0\rangle$ is uplifted leading to three, three and two separated components, respectively. The probabilities along vertical axis are the amplitudes $|c^{g}_{S,M}|^{2}$.}
\label{fig:sntest} 
\end{figure}

The corresponding circuit is shown in Fig.~\ref{fig:circuit_sn}.  This circuit is implemented to perform calculations utilizing qiskit \cite{Abr19}, and the results obtained for the same condition as in panel (b) of Fig.~\ref{fig:testsimple} are shown in Fig.~\ref{fig:sntest}. We see in this figure that the amplitudes associated previously with the two components $| \Psi_{1,M} \rangle$ with $M=-1,1$ have now systematically split into 
 three amplitudes corresponding to the three states $| {1,M} \rangle_{g=1,2,3}$.  Similarly, the component associated to $| \Psi_{0,0} \rangle$
 is now separated into the two contributions $| 0,0\rangle_0$ and  $| 0,0\rangle_1$. Here, again, the algorithm has been validated by confronting the amplitudes obtained numerically with the analytical ones. Finally, we mention that the outcome of the circuit after each measurement is one of the states of the irreducible total spin 
 representation.    
    
\subsection{Reducing the circuit depth of the TQSf method}

The brute-force generalization of the algorithm to project a given state onto one of the irreducible representations of the total spin 
requires a rather larger number of operations and of ancillary qubits. As seen in Eq.~\eqref{eq:transpos}, the number of transpositions in ${\bf S}^2_{[j]}$
is equal to $j(j-1)/2$. Therefore, if the Trotter-Suzuki technique is employed to simulate the operator $U_{[j]}$, the exponential appearing 
in this operator have {\it a priori} also to be split  into $j(j-1)/2$ terms. 
To reduce the numerical efforts, first, we note that the states $|S, M \rangle_g$ are also eigenstates of the difference ${\bf S}^2_{[j]} - {\bf S}^2_{[j-1]}$ for $2 \ge j \ge n$. Since we have
 \begin{eqnarray}
  {\bf S}^2_{[j]} -   {\bf S}^2_{[j-1]} &=&  \frac{5 -2j}{4}  + \sum_{i < j} P_{ij},  \label{eq:dsh}
 \end{eqnarray}
 we finally remark that these states are the eigenstates of the set of simpler operators given by
 \begin{eqnarray}
 H_{[j]} &=& \sum_{i < j} P_{ij} \label{eq:hj},
\end{eqnarray} 
for $j=2,\cdots, n$. The eigenvalues of a given operator $H_{[j]}$ are integers  and lie in the interval $[-1, j-1]$. 
The set of eigenvalues of the $H_{[j]}$ also uniquely defines a state of the irreducible representation. If we denote an eigenvalue  
of $H_{[j]}$ by $h_{[j]}$,  
the three different paths displayed in Eq.~\eqref{eq:paths} correspond to the sequences $[ h_{[2]} \rightarrow h_{[3]} \rightarrow h_{[4]} ]$
respectively given by (a) $[ +1 \rightarrow  +2 \rightarrow -1]$, (b) $[ +1 \rightarrow  -1 \rightarrow +2]$ and (c) $[ -1 \rightarrow +1 \rightarrow +2]$. Therefore, they can  be used as an alternative of the ${\bf S}^2_{[j]}$ in the previous algorithm. A proper choice of the $U_{[n]}$ is then:
\begin{eqnarray}
U_{[n]} &=& \exp\left\{ 2\pi i \frac{\left[ H_{[j]} + 1 \right]} {2^{n_{[j]}}}\right\},
\end{eqnarray}     
where $n_{[j]}$ is optimally chosen as the minimal value of $n_{[j]}$ verifying for $j\ge 2$
\begin{eqnarray}
n_{[j]} &>& \frac{\ln(j-1)}{\ln 2}. \label{eq:njh}
\end{eqnarray} 
The use of $H_{[j]}$ instead of ${\bf S}^2_{[j]}$ has two practical advantages.  As seen from Eq.~\eqref{eq:hj}, these operators contain only 
$(j-1)$ transpositions, and therefore the number of terms in the Trotter-Suzuki method will scale linearly with $j$ compared to the quadratic number of terms for ${\bf S}^2_{[j]}$. In addition, the number of ancillary qubits $n_{[j]}$ obtained from the condition given in Eq.~\eqref{eq:njh} will also be much lower than the one obtained from the previous condition given in Eq.~\eqref{eq:nsoddeven}  when $j$ increases.  We have also implemented 
the TQSf approach based on the operators $\{ H_{[j]} \}$ for the illustration given in Fig.~\ref{fig:sntest} and have 
obtained strictly the same results (not shown here) but with a less number of operations. 
  
\subsection{TQSf method based on sequential measurements technique with minimal quantum resources}
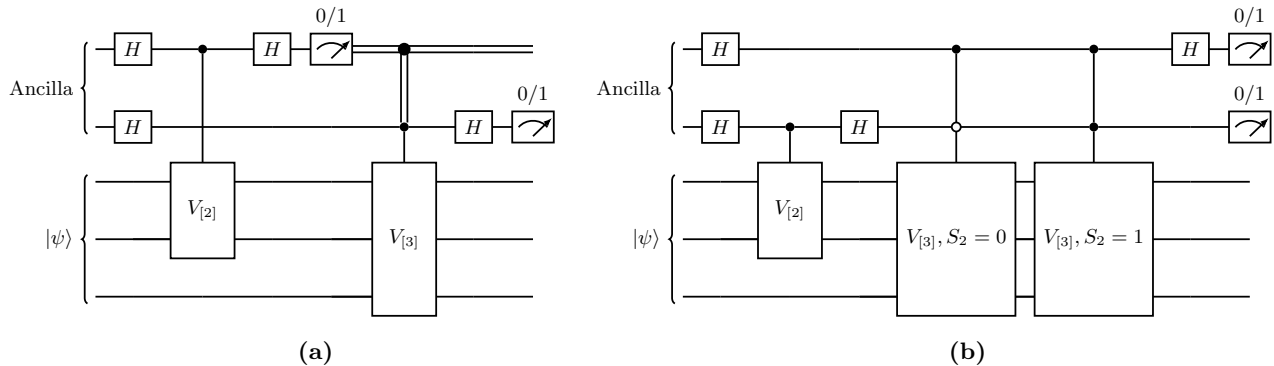
\begin{figure*}
	\begin{center}
		\begin{tikzpicture}
			\node[scale=0.85] {
				\begin{quantikz}[row sep=0.3cm,column sep =0.3 cm]
					\lstick[wires=2]{Ancilla}&\gate{H}& \ctrl{2} & \gate{H}&\meter{0/1}&\cwbend{1}& \cw& \cw \\
					&\gate{H}& \qw& \qw & \qw & \ctrl{1} &\gate{H} & \meter{0/1}\\
					\lstick[wires=3]{$|\psi\rangle$}&\qw & \gate[wires=2][1cm]{V_{[2]}}& \qw & \qw&\gate[wires=3][1cm]{V_{[3]}} & \qw& \qw \\
					&\qw& \qw& \qw & \qw & \qw & \qw& \qw \\
					&\qw& \qw& \qw & \qw & \qw & \qw& \qw 
				\end{quantikz}
			};
			\node at (0.5,-2.5) [scale=1]{\textcolor{black}{\textbf{(a)}}};
		\end{tikzpicture}
		\begin{tikzpicture}
			\node[scale=0.85] {
				\begin{quantikz}[row sep=0.3cm,column sep =0.3cm]
					\lstick[wires=2]{Ancilla}&\gate{H}& \qw& \qw & \ctrl{1} & \ctrl{1}&\gate{H} & \meter{0/1}\\
					&\gate{H}& \ctrl{1} & \gate{H}&\octrl{1} &\ctrl{1}& \qw& \meter{0/1}\\
					\lstick[wires=3]{$|\psi\rangle$}&\qw& \gate[wires=2][1cm]{V_{[2]}}& \qw&\gate[wires=3][1cm]{V_{[3]},S_{2}=0}&\gate[wires=3][1cm]{V_{[3]},S_{2}=1} & \qw& \qw \\
					&\qw& \qw& \qw & \qw & \qw & \qw& \qw\\
					&\qw& \qw& \qw & \qw & \qw & \qw& \qw
				\end{quantikz}
			};
			\node at (0.5,-2.5) [scale=1]{\textcolor{black}{\textbf{(b)}}};
		\end{tikzpicture}
		\caption{Quantum circuits for a three spin system to implement the technique for Young tableaux to encode the path of total spin. (a) The circuit with the controlled operations on classical bits based on the measurement outcome from previous step. Double lines represent the classical bits. (b) An alternative circuit based on the principle of deferred measurement~\cite{Nie02}, suitable for currently available real quantum hardware. These circuits can be extended for a larger spin system in a similar manner.}\label{fig:dkcircuit}
	\end{center}
\end{figure*}
In the previous discussion, we have explored the possibility of obtaining the amplitudes of any state on the total spin basis 
by performing the simultaneous measurements of a set of ancillary qubits. These measurements give a snapshot of the 
paths of each total spin eigenvectors in the so-called sequential construction of the state.  

As underlined in Ref.~\cite{Jor10} and further discussed recently in Refs.~\cite{Hav18,Hav19}, one can associate a binary number to each path representing directly the increase or decrease of the total spin components or Young tableaux construction (see Fig.~4 of Ref.~\cite{Hav19}). Thus, considering the three examples of paths in Eq.~\eqref{eq:paths} again, the different paths can indeed be represented by (a) $[ \nearrow~\nearrow ~\searrow]$, (b)    $[ \nearrow  ~\searrow ~ \nearrow]$   and (c) $[ \searrow ~ \nearrow ~\nearrow]$ that can be associated with the three binary numbers $110$, $101$ and $011$, respectively. 

A possible manner to directly encode the increase or decrease of the total spin on a single qubit is to find an appropriate 
operator to encode this property. Let us assume that we have $j-1$ qubits already having a total spin $S_{[j-1]}$ 
that is known. If we add one more spin, the new total spin that is accessible to the complete set of spins 
will be $S_{[j]} =  S_{[j-1]} \pm 1/2$  (note that $S_{[j-1]} = 0$ imposes
$S_{[j]} =  S_{[j-1]} + 1/2$).  A simple analysis shows that the following operator
\begin{eqnarray}
G_{[j]} &=&  \frac{{\bf S}^2_{[j]}-{\bf S}^2_{[j-1]} +  S_{[j-1]} + \frac{1}{4}  }{(2 S_{[j-1]} + 1)}, \label{eq:gj}
\end{eqnarray}   
has an eigenstate of the total spin with an eigenvalue equal to $1$ (resp. $0$) for the eigenstate associated 
to the spin $ S_{[j]} = S_{[j-1]} + 1/2$ (resp. $S_{[j]} = S_{[j-1]} - 1/2$).  Therefore, this operator directly encodes 
the increase or decrease of the total spin when adding the spin $j$.  

An important aspect of the application of the operator $G_{[j]}$ is that (i) the mapping to a single binary digit is valid only if the 
set of $j-1$ spins are already projected onto eigenstates of the total spin ${\bf S}^2_{[j-1]}$ and (ii) the eigenvalue $S_{[j-1]}$  is known.  
Assuming that these two conditions are fulfilled, it is worth mentioning that a single ancillary qubit will be necessary to perform the QPE 
method for the $G_{[j]}$ operator. The unitary operator to be used in the QPE is given by 
\begin{eqnarray}
V_{[j]} &=& \exp\{\pi i G_{[j]} \},
\end{eqnarray}   
and the QPE reduces to a simple Hadamard test.  

The two conditions above suggest a modified algorithm with an iterative procedure for the measurements with a successive set of 
projections on the ${\bf S}^2_{[j]}$ with increasing $j$. We restart from a system $| \Psi \rangle$ described on a set of $n$ spins.  We introduce 
a variable $S$ that will be updated at each measurement and equal to the $S_{[j]}$ value at step $j$. Initially, $S=1/2$, i.e., the total 
spin of a single spin.  Consecutively, we make the set
of Hadamard tests/measurements iteratively as follows\\
\\
\hspace*{1.0cm} $S=\frac{1}{2}, j=1$ \\
\hspace*{1.0cm} while $j \neq n$  do\\
\hspace*{1.5cm} $j \rightarrow j+1$   \\
\hspace*{1.5cm} if  $S \neq 0$ do   \\
\hspace*{2.0cm} $S_{[j]} = S$    \\
\hspace*{2.0cm} Perform the Hadamard test with $V_{[j]}$ \\
\hspace*{2.0cm} Measure the ancillary qubit  \\
\hspace*{2.0cm} $M =$ result of the measurement ($0$ or $1$)   \\
\hspace*{2.0cm} $S \rightarrow S + M - \frac{1}{2}$ \\
\hspace*{1.5cm} else do   \\
\hspace*{2.0cm} $S \rightarrow S+\frac{1}{2}$     \\
\hspace*{1.5cm} end if   \\
\hspace*{1.0cm}  end while \\
\hspace*{1.0cm} $S_{[n]} = S$    \\

One difficulty in the algorithm is that the intermediate step $j$ is triggered by the knowledge of $S_{[j-1]}$ and more generally of the 
total spins components $S_{[k]}$ with $k < j$. Assuming ideally that the interface from a quantum to a classical computer works perfectly, 
the above algorithm can be implemented using sequentially a set of Hadamard tests for the operators $V_{[k]}$ with increasing $k$.  
Explicitly, starting from the initial state $| \Psi \rangle$, an Hadamard test is performed using $V_{[2]}$, after the ancillary qubit measurement, the value of $S$ is updated on the classical computer and the new operator $V_{[3]}$ is constructed. Then, a second Hadamard test is made 
on the system using $V_{[3]}$, and so on, until all Hadamard tests are performed. This procedure is nothing but a quantum algorithm 
with repeated controlled operations by the classical computer. 
    
We show in Fig.~\ref{fig:dkcircuit} the circuit for a three spin case to perform this scheme (panel (a)). We start with considering two spins, and after measurement, the resulting value ($0$ or $1$) stored in the classical bit is utilized to control the form of operator $V_{[j]}$ to be considered for the three spin case. Furthermore, by adding one more ancilla qubit for each spin, and the controlled operations using the values stored previously on classical bits, we can extend this circuit to explore the larger spin/qubit systems. 

\begin{figure}[h!]
	\includegraphics[width=0.99\linewidth]{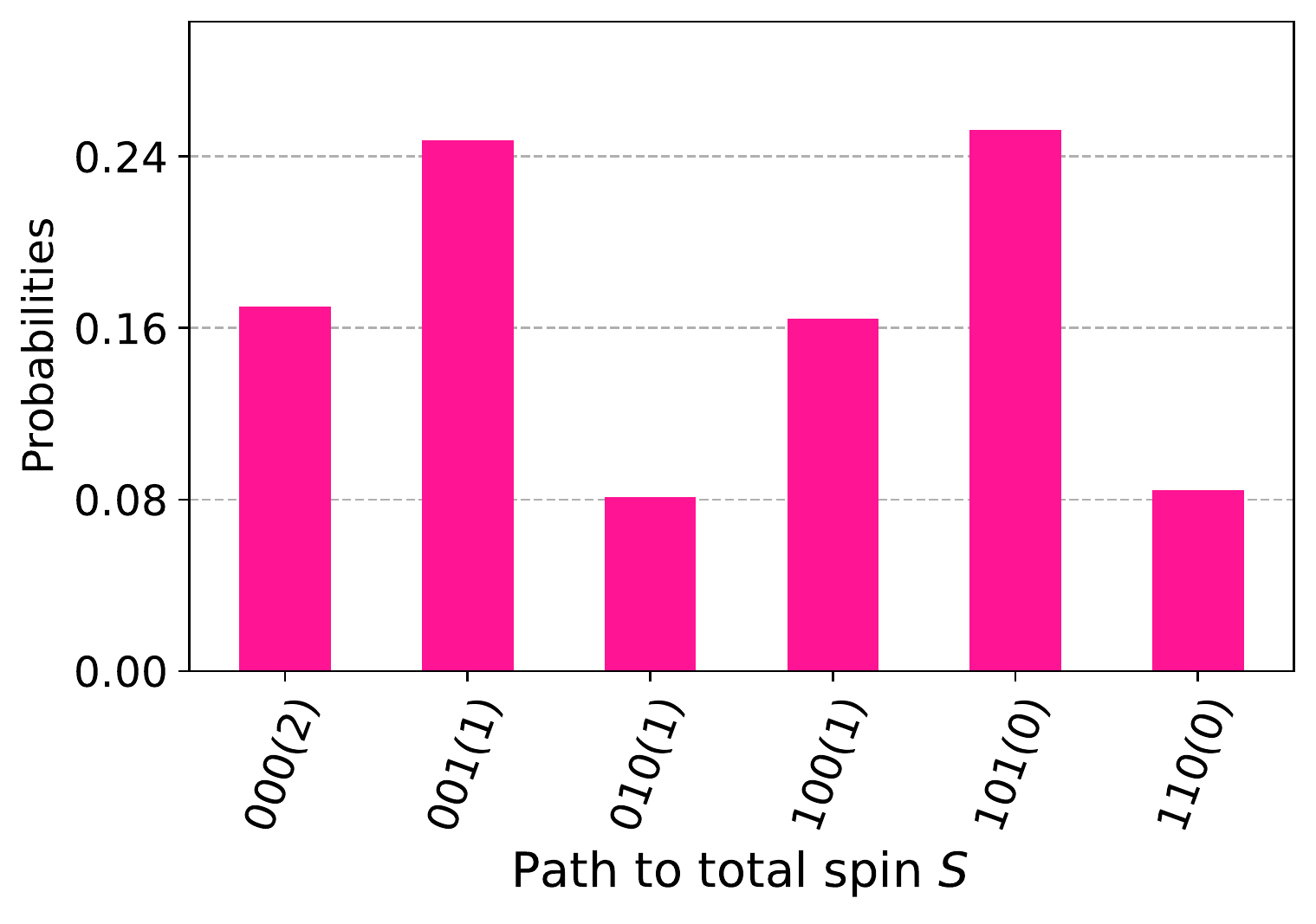}
	\caption{Illustration of the results obtained for a system described on $n=4$ qubits with the same initial state as in panel (b) of Fig. \ref{fig:testsimple} but using the circuit shown in Fig. \ref{fig:dkcircuit} (b). The $0 ~(1)$ in the bit strings on horizontal axis represents the increase (decrease) in the spin, and the total spin $S$ is given in parenthesis. The path represented by the bit strings should be read from right to left.}
	\label{fig:dk} 
\end{figure}

Considering the same initial state as used in Fig.~\ref{fig:testsimple} (b), the results obtained from the extension of circuit Fig.~\ref{fig:dkcircuit} (a) to four qubit/spin case are given in Fig.~\ref{fig:dk}. As a straightforward validation, we can see that the contribution of total $S$ is the same as given in Fig.~\ref{fig:sntest}. To obtain the irreducible representation, we can project the $S_{z}$ in the same way as performed in the earlier two techniques.

 We finally mention that conditional operators on the classical register are currently not supported on the available real quantum devices.  Therefore, we also explore the possibility to apply the present procedure without requesting classical controlled operations.   
 An alternative procedure is to use the circuit given in Fig.~\ref{fig:dkcircuit} (b), which is based on the principle of deferred measurement~\cite{Nie02}. This procedure for the three qubit case essentially leads to the same results. But this circuit has complexities in the form of multi-controlled gates, which need to be further decomposed into single and two-qubit gates.

\section{Conclusions}

Starting from a general wave function described on a set of spins/qubits, we address the problem of its decomposition onto 
eigenstates of the total spin. We begin with the methodology proposed in Ref.~\cite{Lac20} to obtain a minimal of quantum algorithms that play the role of projection onto total angular momentum and total spin 
azimuthal projection. The different algorithms have various degrees of sophistication depending on 
the requested tasks. They can either solely project onto the subspace with good $S$ and $M$ components or completely
uplift the degeneracy in this subspace. The measurement of the ancillary qubits gives access to the amplitudes of the initial states 
on a total spin basis. After each measurement, the state collapses to one of the eigenstates of the total spin. Therefore, the procedure can be used as a filter to prepare such eigenstates. For this reason, we call the method Total Quantum Spin filtering (TQSf). We propose several methods, either performing the operations on a quantum computer only or mixing quantum-classical computation.  In the latter case, the quantum resources are minimized.    

The method can have a wide 
range of applications. The first one, the original motivation of the present work, is associated with the spin symmetry restoration 
in many-body systems such as those appearing in quantum chemistry, nuclear physics, or condensed matter physics. In this case, a parametrized spin
symmetry breaking state can be used as the initial state, and the TQSf can be performed prior to the state parameter optimization. This variation after 
projection method \cite{Rin80,Bla86} is known to be rather effective but states obtained in this way are difficult to manipulate on a classical computer.  

We mention in the article that the method can also be used to generate random numbers on a discretized mesh. 
This aspect could be further explored in the future by connecting our study with spin random walk theory \cite{Kem09}.
Another avenue that could be interesting to explore is the possibility to generalize the present method to construct tensor 
networks (see, for instance, Ref.~\cite{Tic20} for the recent discussion on the connection between the SU(2) algebra and tensor networks).
               
\section*{Acknowledgments}

This project has received financial support from the CNRS through
the 80Prime program and is part of the QC2I-IN2P3 project. We acknowledge
the use of IBM Q cloud as well as use of the Qiskit software package
\cite{Abr19} for performing the quantum simulations.

\end{document}